\documentclass[12pt]{article}
\pdfoutput = 1
\usepackage{amsmath,amssymb,bm,epsfig,afterpage}
\usepackage{mathtools}
\usepackage{cite}
\usepackage{here}
\usepackage{color}
\usepackage{tabularx}
\usepackage{simplewick} 
\usepackage{bbm}

\usepackage{comment}

\renewcommand{\thefootnote}{\fnsymbol{footnote}}
\newcommand{\vev}[1]{{\langle{#1}\rangle}}

\newcommand{\abs}[1]{\left|{#1}\right|}

\newcommand{\eps}{\epsilon}

\newcommand{\order}[1]{\mathcal{O}\left({#1}\right)}
\newcommand{\Ykr}[2]{Y^{({#1})}_{#2}} 

\newcommand{\pr}{\prime}
\newcommand{\ppr}{{\prime\prime}}

\newcommand{\Arg}{\mathrm{Arg}}

\newcommand{\MeV}{\mathrm{MeV}}
\newcommand{\GeV}{\mathrm{GeV}}
\newcommand{\TeV}{\mathrm{TeV}}
\newcommand{\CW}{{\mathrm{CW}}}

\newcommand{\ita}{\mathrm{Im}\,\tau}
\newcommand{\rta}{\mathrm{Re}\,\tau}

\numberwithin{equation}{section}

\allowdisplaybreaks[3]
\newcolumntype{Y}{&gt;{\centering\arraybackslash}X} 

\usepackage{hyperref}

\definecolor{asparagus}{rgb}{0.53, 0.66, 0.42}
\definecolor{darkspringgreen}{rgb}{0.09, 0.45, 0.27}
\definecolor{darkturquoise}{rgb}{0.0, 0.81, 0.82}
\definecolor{dollarbill}{rgb}{0.52, 0.73, 0.4}

\begin{document}

\begin{titlepage}

\begin{flushright}
 {\tt
CTPU-PTC-24-02 \\
EPHOU-24-001  
}
\end{flushright}

\vspace{1.2cm}
\begin{center}
{\Large
{\bf
Finite modular axion and radiative moduli stabilization 
}
}
\vskip 2cm
Tetsutaro Higaki$^{\ a}$~\footnote{thigaki@rk.phys.keio.ac.jp} 
and 
Junichiro Kawamura$^{\ b}$~\footnote{junkmura13@gmail.com}
and
Tatsuo Kobayashi$^{\ c}$~\footnote{kobayashi@particle.sci.hokudai.ac.jp}

\vskip 0.5cm

{\it $^a$
Department of Physics, Keio University, Yokohama, 223-8522, Japan
}\\[3pt]

{\it $^b$
Center for Theoretical Physics of the Universe, Institute for Basic Science (IBS),
Daejeon 34051, Korea
}\\[3pt]

{\it $^c$
Department of Physics, Hokkaido University, Sapporo 060-0810, Japan}\\[3pt]

\vskip 1.5cm

\begin{abstract}
We propose a simple setup which can stabilize a modulus field 
of the finite modular symmetry by the Coleman-Weinberg potential.
Our scenario leads to  a large hierarchy 
suppressing instanton-like corrections $e^{2\pi i\tau}$ 
and to a light axion identified as $\mathrm{Re} \tau$, 
where $\tau$ is the modulus field.
This stabilization mechanism provides 
the axion solution to the strong CP problem.  
The potential has a minimum at a large $\mathrm{Im}\tau$ 
which suppresses explicit $U(1)_{\mathrm{PQ}}$ violation terms 
proportional to $e^{-2\pi {\mathrm{Im}\tau}}$, 
and hence the quality of the axion is ensured 
by the residual symmetry associated with the $T$-transformation, 
$\tau \to \tau +1$,
around the fixed point $\tau \sim i\infty$.   
\end{abstract}
\end{center}
\end{titlepage}

\clearpage

\renewcommand{\thefootnote}{\arabic{footnote}}
\setcounter{footnote}{0}

\newcommand{\la}{{\lambda}}
\newcommand{\ka}{{\kappa}}
\newcommand{\mQ}{{m^2_{\tilde{Q}}}}
\newcommand{\mU}{{m^2_{\tilde{u}}}}
\newcommand{\mD}{{m^2_{\tilde{d}}}}
\newcommand{\mL}{{m^2_{\tilde{L}}}}
\newcommand{\mE}{{m^2_{\tilde{e}}}}
\newcommand{\mhu}{{m^2_{H_u}}}
\newcommand{\mhd}{{m^2_{H_d}}}
\newcommand{\ms}{{m^2_S}} 
\newcommand{\Ala}{{A_\lambda}}
\newcommand{\Aka}{{A_\kappa}} 

\newcommand{\id}[1]{\mathbf{1}_{#1}}
\newcommand{\ol}[1]{\overline{#1}}
\newcommand{\Lcal}{\mathcal{L}}
\newcommand{\Mcal}{\mathcal{M}}
\newcommand{\Ncal}{\mathcal{N}}
\newcommand{\Ycal}{\mathcal{Y}}
\newcommand{\Wcal}{\mathcal{W}} 
\newcommand{\sg}{\sigma}
\newcommand{\sgb}{\overline{\sigma}}
\newcommand{\del}{\partial}

\newcommand{\htm}{\hat{m}}
\newcommand{\tU}{\widetilde{U}}
\newcommand{\SL}[2]{\mathrm{SL}({#1},{#2})}
\newcommand{\natN}{\mathbb{N}}
\newcommand{\intZ}{\mathbb{Z}}
\newcommand{\oGam}{\overline{\Gamma}} 
\newcommand{\Gam}{\Gamma} 
\newcommand{\Ita}{\mathrm{Im}\,\tau}

\newcommand{\QCD}{\mathrm{QCD}}
\newcommand{\CKM}{\mathrm{CKM}}
\newcommand{\SM}{\mathrm{SM}}
\newcommand{\hY}{\hat{Y}} 

\newcommand{\Acal}{\mathcal{A}}
\newcommand{\tM}{\tilde{M}}

\newcommand{\nop}[1]{\textcolor{blue}{#1}}
\newcommand{\LQCD}{\Lambda_{\mathrm{QCD}}}

\newcommand{\YA}[1]{{\color{darkviolet} [\textbf{YA:} #1]}}
\definecolor{darkviolet}{rgb}{0.58, 0.0, 0.83}
\newcommand{\violed}[1]{{\color{darkviolet}#1}}
\newcommand{\magenta}[1]{{\color{magenta}#1}}

\newcommand{\PQ}{\mathrm{PQ}}
\newcommand{\MSbar}{\overline{\mathrm{MS}}}

\section{Introduction}

The finite modular symmetry has been intensively 
studied to explain the flavor structure 
of the Standard Model (SM) quarks and leptons~\cite{Feruglio:2017spp,Kobayashi:2018vbk,Penedo:2018nmg,Novichkov:2018nkm,Ding:2019xna,Liu:2019khw,Novichkov:2020eep,Liu:2020akv,Liu:2020msy}. 
(See for reviews~\cite{Kobayashi:2023zzc,Ding:2023htn}.)
Under the symmetry, a Yukawa coupling constant behaves as a so-called modular form  
which is a holomorphic function of the modulus $\tau$. 
Interestingly, the finite modular symmetries $\Gamma_N$, $N\in\natN$, 
are isomorphic to the non-Abelian discrete symmetries~\cite{deAdelhartToorop:2011re}, 
which can explain the flavor structure of the SM fermions~\cite{Altarelli:2010gt,Ishimori:2010au,Hernandez:2012ra,King:2013eh,King:2014nza,Tanimoto:2015nfa,King:2017guk,Petcov:2017ggy,Feruglio:2019ybq,Kobayashi:2022moq}. 
Such modular dependence of the Yukawa coupling to a modulus 
was found in heterotic orbifold models~\cite{Lauer:1989ax,Lauer:1990tm,Ferrara:1989qb,Baur:2019kwi,Nilles:2020nnc,Nilles:2020gvu} 
and magnetized D-brane models~\cite{Kobayashi:2017dyu,Kobayashi:2018rad,Kobayashi:2018bff,Ohki:2020bpo,Kikuchi:2020frp,Kikuchi:2020nxn,Hoshiya:2020hki} 
in superstring theory. 
Moreover, the finite modular symmetry 
can explain the hierarchies of the quark and lepton masses and mixing 
if the modulus is stabilized near one of the fixed points 
where the residual symmetry remains unbroken~\cite{
Petcov:2022fjf,Petcov:2023vws,Kikuchi:2023jap, Abe:2023ilq,Kikuchi:2023cap,
Feruglio:2021dte,Novichkov:2021evw,Abe:2023qmr,Abe:2023dvr,deMedeirosVarzielas:2023crv,
Kikuchi:2023dow,Kikuchi:2023fpl}.

In the models with finite modular flavor symmetries, 
the value of modulus $\tau$ plays an essential role to fit to the data   
and is often treated as a free parameter.  
Whereas in a UV theory, its value should be fixed dynamically 
through modulus stabilization. 
There are studies on the moduli stabilization 
with the three-form fluxes~\cite{Ishiguro:2020tmo,Ishiguro:2022pde},  
the negative power of modular form~\cite{Kobayashi:2019xvz,Kobayashi:2019uyt}~\footnote{
The positive power of the trivial singlet is also studied in Ref.~\cite{Kobayashi:2019uyt}. See also Ref.~\cite{Abe:2023ylh}.
}, 
and the general superpotential invariant under the modular symmetry $SL(2,\intZ)$~\cite{Novichkov:2022wvg,Leedom:2022zdm,Knapp-Perez:2023nty,Feruglio:2023uof,King:2023snq}, 
utilizing the Klein $j$ function~\cite{Cvetic:1991qm}. 
The latter two mechanisms will need non-perturbative dynamics 
to generate those potentials. 
Recently, effect of the radiative correction to the tree-level stabilization 
is discussed in Ref.~\cite{Kobayashi:2023spx}.

In this work, we point out that the modulus can be stabilized 
only by the Coleman-Weinberg (CW) potential generated by couplings with matter fields. 
This mechanism provides a perturbative way to stabilize a modulus
through the supersymmetry breaking. 
For illustration, we shall consider a simple model 
with vector-like quarks which transform as non-trivial singlets 
under the finite modular flavor symmetry. 
The resultant potential has a global minimum at $\ita \gg 1$, 
where the residual $\intZ^T_N$ symmetry associated with the $T$ transformation, 
$\tau \to \tau+1$, is unbroken.  
That can lead to a large hierarchy suppressing instanton-like correction, i.e., $e^{-2\pi {\ita}} \ll 1$. 
An interesting feature of this mechanism 
is that the $\rta$ direction can be identified as the QCD axion 
due to 
the accidental Peccei-Quinn (PQ) symmetry 
originated from the $\intZ^T_N$ residual symmetry. 
We shall discuss the condition 
that the axion solution to the strong CP problem is not spoiled 
due to the CW potential, i.e. the axion quality is good enough.

This paper is organized as follows. 
The finite modular symmetry is briefly reviewed in Sec.~\ref{sec-fmod}. 
In Sec.~\ref{sec-stab}, we discuss the mechanism of the modular stabilization 
by utilizing the CW potential,
and then we argue existence of the QCD axion in the setup in Sec.~\ref{sec-axion}.
Section~\ref{sec-concl} concludes.

\section{Finite modular symmetry}
\label{sec-fmod}

We consider the series of groups  
\begin{align}
 \Gamma(N) :=  \left\{ 
\begin{pmatrix}
 a & b \\ c & d 
\end{pmatrix}
\in \mathrm{SL}(2,\intZ), 
\quad 
\begin{pmatrix}
 a & b \\ c & d 
\end{pmatrix}
 \equiv 
\begin{pmatrix}
1 & 0 \\ 0 & 1 
\end{pmatrix}
\quad 
\mathrm{mod}~N
\right\},
\end{align}
where $N \in \natN$ is called level 
and $\Gamma := SL(2,\intZ) = \Gamma(1)$ 
is a group of $2\times 2$ matrices with determinant unity. 
The modular transformation $\gamma \in \Gamma(N)$ for a complex parameter $\tau$ 
is defined as 
\begin{align}
 \tau \to \gamma\tau = \frac{a\tau+b}{c\tau+d}. 
\end{align}
The modular group is generated by three generators 
\begin{align}
 S = 
\begin{pmatrix}
 0 & 1 \\ -1 & 0
\end{pmatrix}, 
\quad 
T = 
\begin{pmatrix}
 1 & 1 \\ 0 & 1
\end{pmatrix}, 
\quad 
R =  
\begin{pmatrix}
 -1 & 0 \\ 0 & -1
\end{pmatrix}.  
\end{align}
The finite modular symmetry is defined as the quotient group   
$\Gamma_N := \ol{\Gamma}/\Gamma(N)$ where $\ol{\Gamma} := \Gamma(1)/\intZ^R_2$. 
Here $\intZ^R_2$ is the $\intZ_2$ symmetry generated by $R$. 
Under $\Gamma_N$ with $N < 6$, the generators satisfy 
\begin{align}
S^2 = (ST)^3 =  T^N = 1. 
\end{align}
There are Abelian discrete symmetries 
$\intZ^S_2$, $\intZ^{ST}_3$ and $\intZ^T_N$ in associated 
with the $S$, $ST$ and $T$ generators 
which are unbroken at $\tau = i$, $w := e^{2\pi i/3}$ and $i\infty$, respectively. 
The finite modular symmetries are isomorphic to the non-Abelian discrete symmetries, 
e.g. $\Gamma_3 \simeq A_4$.  
A modular form $\Ykr{k}{r}(\tau)$ with a modular weight $k$ and representation $r$ 
transforms under $\Gamma_N$ as 
\begin{align}
 \Ykr{k}{r}(\tau) \to (c\tau+d)^k \rho(r) \Ykr{k}{r}(\tau),
\end{align}
where $\rho(r)$ is the representation matrix of $\Gamma_N$. 
We assume that the chiral superfield $Q$ with weight $-k_Q$ and representation $r_Q$
transforms as 
\begin{align}
 Q \to (c\tau+d)^{-k_Q} \rho(r_Q) Q. 
\end{align}
More detailed discussions can be found in e.g. Refs.~\cite{Kobayashi:2023zzc,Ding:2023htn}.

\section{Radiative moduli stabilization}
\label{sec-stab}

We consider a simple supersymmetric model with the following 
K\"ahler potential and superpotential:  
\begin{align}
 K =&\ -h \log(-i\tau + i\tau^\dag) 
     + \sum_i \left(
         \frac{Q_i^\dag Q_i}{(-i\tau+i\tau^\dag)^{k_{Q_i}}}  
        +\frac{\ol{Q}_i^\dag \ol{Q}_i}{(-i\tau+i\tau^\dag)^{k_{\ol{Q}_i}}}  
     \right), 
\\ \notag 
W =&\ \sum_{i} M_{Q_i} \Ykr{k_{i}}{r_{i}}(\tau) \ol{Q}_i Q_i,   
\end{align}
where $h\in \natN$. 
The reduced Planck mass $M_p=2.4\times 10^{18}~\GeV$ is set to unity.
The chiral superfield $Q_i$ ($\ol{Q}_i$) 
has a modular weight $k_{Q_i}$ ($k_{\ol{Q}_i}$) 
and is assumed to be a singlet under $\Gamma_N$ for simplicity. 
Here, $Q_i$ and $\ol{Q}_i$ are vector-like pairs under the SM gauge group.   
The weight of the modular form $k_{i} = k_{Q_i} + k_{\ol{Q}_i}- h$  
and the representation $r_i$ are chosen 
so that the combination $e^K |W|^2$ in supergravity  
is invariant under $\Gamma_N$. 
We introduce the vector-like mass parameter $M_{Q_i}$ which could be replaced 
by vacuum expectation values of fields. 
Throughout this work, we assume that the SM fields are trivial singlets 
under the modular symmetry $\Gamma_N$ for simplicity 
and they have no effects in the following analysis.

We shall show that the modulus $\tau$ can be stabilized 
without introducing a superpotential for the modulus $\tau$, 
but with the 1-loop CW potential 
\begin{align}
\label{eq-VCW}
 V_\CW =  \frac{1}{32\pi^2} \sum_i & \left[ 
 \left(m_{i}^2 + m_{Q_i}^2(\tau)\right)^2 
 \left(\log\left(\frac{m_{i}^2+m_{Q_i}^2(\tau)}{\mu^2}\right)-\frac{3}{2}\right)  
\right. \notag  \\ 
& \hspace{3.0cm}\left.  - 
(m_{Q_i}^2(\tau))^2  \left(\log\left(\frac{m_{Q_i}^2(\tau)}{\mu^2}\right)-\frac{3}{2}\right) 
  \right], 
\end{align}
with
\begin{align}
\label{eq-mQi}
 m_{Q_i}^2(\tau) 
:= M_{Q_i}^2 \left(-i\tau+i\tau^\dag \right)^{k_i} \abs{\Ykr{k_i}{r_i}(\tau)}^2.   
\end{align} 
Here, we consider the $\MSbar$ scheme with $\mu$ being the renormalization scale.
The first (second) term in the parenthesis corresponds 
to the scalar (fermion) contribution.
The supersymmetric mass $m_{Q_i}$ is multiplied by the factor $(-i\tau+i\tau^\dag)^{k_i/2}$ 
due to the canonical normalization. 
We introduce the soft supersymmetry breaking mass squared $m_i^2$ 
for the scalar component of $Q_i$ 
which is assumed to be independent of $\tau$ for simplicity.
This is the case if supersymmetry is broken 
by a mechanism irrelevant to the modulus $\tau$.
We shall first see the simplest case analytically at large $\ita$, 
and then see the results numerically later. 
Throughout this paper, 
we assume that the mass parameters are common for $Q_i$'s, 
i.e. $M_{Q_i} =: M_Q$, $m_i =: m_0$ and $k_i=: k$
for simplicity. 
Note that the tree-level scalar potential in supergravity 
is irrelevant when $Q_i=\ol{Q}_i =0$.

\subsection{Simplified analysis} 

We argue the model of radiative stabilization of the modulus $\tau$  
where there is only one pair of $(\ol{Q}, Q)$, so we omit the index $i$. 
The singlet modular form of $\Gamma_N$ can be expanded as 
\begin{align}
\label{eq-qexpand}
 \Ykr{k}{1_t}(\tau) = q^{t/N} \sum_{n=0}^\infty c_n q^n,      
\quad 
 q := e^{2\pi i\tau},  
\end{align}
where $r = 1_t$ is a singlet whose charge is $0\le t < N$ 
under the $\intZ^T_N$ symmetry~\footnote{
In the usual notation~\cite{Feruglio:2017spp}, 
$1_0 = 1$, $1_1 = 1^\pr$ and $1_2 = 1^{\ppr}$ for $A_4$.  
}. 
Assuming $\abs{q} = e^{-2\pi \ita} \ll 1$ and $m_0^2 \ll M_Q^2$, 
the CW potential is  given by 
\begin{align}
 V_\CW = \frac{m_0^2 \tM_Q^2}{16\pi^2} x^k e^{-px} 
 \left( \log\frac{\tM_Q^2}{e \mu^2} + k \log x - px \right) 
 + \order{\abs{q}, m_0^4}, 
\end{align}    
where $x := 2\ita$, $p := 2\pi t/N$ and $\tM_Q := c_0 M_Q$. 
At the leading order, there is no potential for $\rta$ 
which will be important for the axion interpretation discussed in the next section. 
The first derivative of the potential is given by 
\begin{align}
\label{eq-dVCW}
  \frac{dV_\CW}{dx} \sim 
  \frac{m_0^2 \tM_Q^2}{16\pi^2} x^{k-1} e^{-px} \left( k-px \right)
   \left( \log\frac{\tM_Q^2}{\mu^2} + k \log x - px \right).  
\end{align}
The point $x=k/p$ is a maximum for sufficiently large $k/p$,  
and hence this potential can have a minimum at
\begin{align}
\label{eq-xLO}
 x_0 = -\frac{k}{p} 
       \Wcal\left(-\frac{p}{k}\left(\frac{\mu}{\tM_Q}\right)^{2/k}\right),   
\end{align}  
where the second parenthesis of Eq.~\eqref{eq-dVCW} is vanishing.  
Here, $\Wcal(z)$ is the Lambert function satisfying $\Wcal(z)e^{\Wcal(z)} = z$ 
which has two real values for $-e^{-1} < z < 0$, 
and the larger one can be a solution where our approximation $\abs{q}\ll 1$ is valid. 
Taking $\mu=\tM_Q$ and $p=2\pi/3~$, 
the location of the minimum is $x_0 = 3.9,~7.9,~12$ and $16$ 
for $k=6,8,10$ and $12$, respectively. 
$x_0$ increases slightly by decreasing $\mu$. 
Thus we find that there is the minimum at $\ita \gg 1$ in the CW potential. 
Note that this minimum exists only for $t \ne 0$ 
where the modular form $\Ykr{k}{1_t}$ is a non-trivial singlet. 
For the CW potential from the trivial-singlet modular form, 
the potential increases as $\ita$ increases 
in contrast to that from the non-trivial one,  
and thus there is no minimum at $\ita \gg 1$~\footnote{
The potentials discussed in Refs.~\cite{Novichkov:2022wvg,Leedom:2022zdm,Knapp-Perez:2023nty,Feruglio:2023uof,King:2023snq,Cvetic:1991qm} 
have a similar behavior, that is, $V \to \infty$ for $\ita \to \infty$.}. 
See also the potential in Ref.~\cite{Abe:2023ylh}.
Therefore the dominant CW potential should be generated from 
the non-trivial singlet one for the existence of the minimum 
at $\ita \gg 1$, 
so that the potential tends to approach to zero for a sufficiently large ${\rm Im}\tau$.

\subsection{Numerical analysis} 

\begin{figure}[t]
 \centering
\begin{minipage}[t]{0.48\hsize}
\includegraphics[width=0.95\hsize]{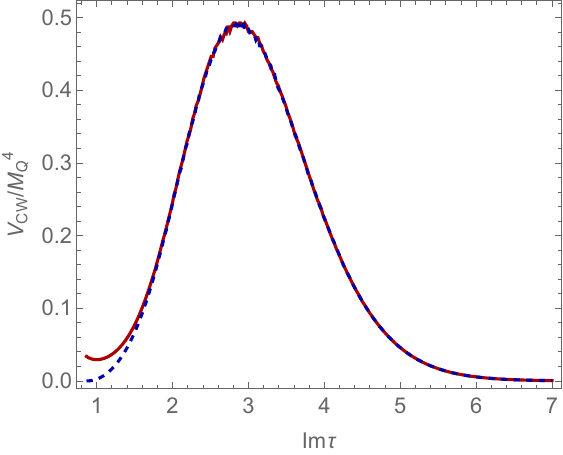}
\end{minipage}
\begin{minipage}[t]{0.48\hsize}
\includegraphics[width=0.95\hsize]{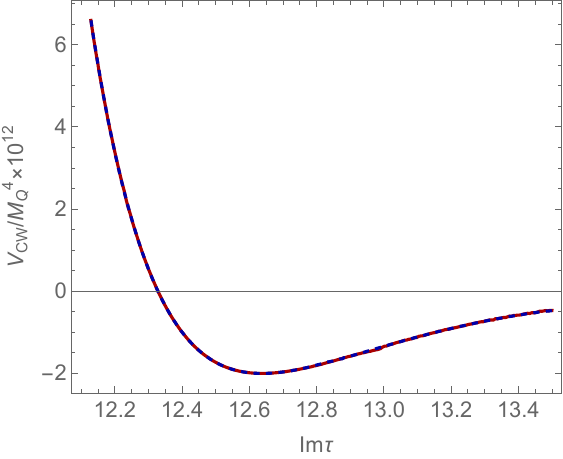}
\end{minipage}
\caption{\label{fig_VCW}
The shape of the CW potential along the $\ita$ direction 
with $\rta = 0$ (red solid) and $\rta=-0.5$ (blue dashed). 
}
\end{figure}

We shall study the potential Eq.~\eqref{eq-VCW} numerically 
without relying on the $q$-expansion in Eq.~\eqref{eq-qexpand}. 
For illustration, we consider $\Gamma_3 \simeq A_4$ as the finite modular group, 
and two pairs of vector-like quarks whose superpotential is given by
\begin{align}
\label{eq-super}
 W = M_Q \sum_{i=1,2} \Ykr{12}{1_i}(\tau) \ol{Q}_i Q_i.  
\end{align}   
We can easily choose the representations and modular weights 
such that only those two terms are allowed but mixing terms like $\bar Q_1 Q_2$ are forbidden.
An explicit assignment of representations and modular weights 
will be shown in Eq.~\eqref{eq-SampleAssign} 
when we shall discuss the vector-like quark decays.  
The modular forms of weight 12 are given by 
\begin{align}
\Ykr{12}{1_1}(\tau) = (Y_1^2 + 2Y_2 Y_3)^2(Y_3^2 + 2Y_1 Y_2), 
\quad 
\Ykr{12}{1_2}(\tau) = (Y_1^2 + 2Y_2 Y_3)(Y_3^2 + 2Y_1 Y_2)^2, 
\end{align}
where the functions $Y_{1,2,3}$ are defined as~\cite{Feruglio:2017spp} 
\begin{align}
\label{eq-YA42}
 Y_1(\tau) =&\ \frac{i}{2\pi} 
        \left[ \frac{\eta^\prime(\tau/3)}{\eta(\tau/3)} 
             + \frac{\eta^\prime((\tau+1)/3)}{\eta((\tau+1)/3)} 
             + \frac{\eta^\prime((\tau+2)/3)}{\eta((\tau+2)/3)} 
            -27\frac{\eta^\prime(3\tau)}{\eta(3\tau)} 
       \right],   \\
 Y_2(\tau) =&\ \frac{-i}{\pi} 
        \left[ \frac{\eta^\prime(\tau/3)}{\eta(\tau/3)} 
             + w^2 \frac{\eta^\prime((\tau+1)/3)}{\eta((\tau+1)/3)} 
             + w \frac{\eta^\prime((\tau+2)/3)}{\eta((\tau+2)/3)} 
       \right],  \\
 Y_3(\tau) =&\ \frac{-i}{\pi} 
        \left[ \frac{\eta^\prime(\tau/3)}{\eta(\tau/3)} 
             + w \frac{\eta^\prime((\tau+1)/3)}{\eta((\tau+1)/3)} 
             + w^2 \frac{\eta^\prime((\tau+2)/3)}{\eta((\tau+2)/3)} 
       \right], 
\end{align}
with $\eta(\tau)$ being the Dedekind eta function. 
The $q$-expansion of the modular forms are given by~\footnote{
There is an ambiguity of the normalization of the modular forms 
which can not be determined from the symmetry.  
We simply employ the normalization used in Ref.~\cite{Feruglio:2017spp}. 
See also Ref.~\cite{Petcov:2023fwh}.
} 
\begin{align}
\label{eq-Y11app}
\Ykr{12}{1_1} = -12 q^{1/3} \left( 1 + 472 q + \order{q^2} \right), 
\quad 
\Ykr{12}{1_2} = 144 q^{2/3} \left( 1 +224 q + \order{q^2} \right).  
\end{align}
The modular weights are chosen to $k=12$ 
so that the strong CP problem is solved 
with keeping the modulus mass heavy, as discussed in the next section~\footnote{
Since $1_1$ and $1_2$ can be in the $2^\pr$ representation under $A_4^\pr$, 
the two modular forms could be embedded into a doublet. 
}. 
Here, we introduce the two vector-like pairs,  
so that the mixed anomaly of $A_4$ with the QCD is canceled~\cite{Araki:2006sqx,Araki:2008ek}. 
The singlets are trivial under the $S$-transformation, 
and transform as $\Ykr{k}{1_t} \to w^t \Ykr{k}{1_t}$ 
under the $T$-transformation. 
Hence, in our setup, $\det(\rho(T)) = w^{1+2} = 1$ and the anomaly is canceled.  
The anomaly associated with the modular weight 
can be canceled by the Green-Schwarz mechanism with an another gauge coupling modulus or by adding
$\log \Ykr{k}{1_0}(\tau)$ to the gauge kinetic function with a certain coefficient, where $\Ykr{k}{1_0}(\tau)$ is a trivial singlet modular form and
could be explained by threshold corrections from heavy modes~\cite{Ibanez:1992hc}.

Figure~\ref{fig_VCW} shows the shape of the potential along the $\ita$ direction. 
Here, we take $m^2/M_Q^2 = 10^{-8}$ and $\mu/M_Q = 10^{-2}$ 
so that the minimum resides at $\ita \simeq 13$.
The red solid (blue dashed) line corresponds to $\rta = 0$ ($-0.5$).  
Note that the fundamental domain is $\ita \ge 1$ ($\sqrt{3}/2$) at $\rta = 0~(-0.5)$.
In the left panel, the origin of the horizontal axis is $\sqrt{3}/2$ and
the global picture of the potential is shown. 
The right panel is the shape of the potential around the global minimum at large $\ita$.
For $\ita \lesssim 1$, where the $q$-expansion is not efficient,   
there is a local minimum at $\tau \sim w$, 
but $V_\CW(\tau=w) = 0$, since $\Ykr{12}{1_{1,2}}(w) = 0$, 
is shallower than the global minimum 
at $\ita \sim 12$ where the potential value is negative.
The potential clearly depends on $\rta$ at small $\ita$, 
while it is independent of $\rta$ within numerical precision. 
Thus, the CW potential in the $\rta$ direction is approximately flat near the global minimum at $\ita \gg 1$.
As a result, our scenario of modulus stabilization can lead a light axion as well as a large hierarchy by $|q| = e^{-2\pi \ita} \sim 10^{-36}$.
We will confirm this flatness analytically in the next section 
to discuss the quality of the axion solution to the strong CP problem.

\section{Axion solution and its quality}
\label{sec-axion}

We have seen that the CW potential Eq.~\eqref{eq-VCW} 
induced by the superpotential Eq.~\eqref{eq-super} 
is very flat in the $\rta$ direction.
The stabilization mechanism can realize the axion solution to the strong CP problem 
by assuming that the matter fields are vector-like quarks as 
in the KSVZ axion model~\cite{Kim:1979if,Shifman:1979if}. 
The effective $\theta$-angle $\ol{\theta}$ in the QCD is given by 
\begin{align}
\label{eq-theta}
\ol{\theta}(\tau) 
= \theta_0 + \Arg \left( \Ykr{12}{1_1}(\tau)\Ykr{12}{1_2}(\tau) \right) 
= \theta_0 + \phi + \order{\abs{q}}, 
\end{align}
where $\theta_0$ is a constant and $\phi := 2\pi\rta$.  
The $\phi$ dependence appears as a result of the 
chiral anomaly of the QCD with the approximate $U(1)_{\rm PQ}$, 
which originates from the residual $\intZ^T_3$ symmetry as shown below.
For $\ita \gg 1$, the Yukawa coupling is given by 
\begin{align}
\Ykr{k}{1_t} \ol{Q}_t Q_t \propto 
e^{ i (t/3) \phi}\; \ol{Q}_t Q_t, 
\end{align}
so we can find an accidental $U(1)_\PQ$ symmetry as 
\begin{align}
\phi \to \phi + \alpha, \quad \ol{Q}_t Q_t \to e^{-i t \alpha/3} \ol{Q}_t Q_t, 
\end{align}
where $\alpha$ is a real transformation parameter. 
Hence, the PQ charge of $\ol{Q}_t Q_t$ is $-t/3$,
and the $U(1)_{\rm PQ}$ is anomalous.
This symmetry is ensured from the residual $\intZ^T_3$ symmetry 
corresponding to $\alpha = 2\pi$, 
and the explicit breaking effects of the $U(1)_\PQ$ 
are from $q = e^{2\pi i\tau}$ which is invariant under $\intZ^T_3$ 
but violates $U(1)_\PQ$~\footnote{
The corrections to the axion from $\log \Ykr{k}{1_0}$ 
in the gauge kinetic function which may exist to cancel the weight anomaly 
will also appear at $\order{q}$. 
 }. 
Therefore, $\rta$ can be identified as the QCD axion. 
Since the kinetic term of the modulus is given by~\footnote{
We write explicitly the reduced Planck mass $M_p$ in this section.  
}
\begin{align}
\label{eq-modkin}
  \frac{h M_p^2}{(2\ita)^2} \partial_\mu \tau^\dag \partial^\mu \tau,   
\end{align}
the axion decay constant is 
$f_a = \sqrt{h} M_p/(2\pi x_0) \sim \order{10^{16}}$ GeV
which requires a fine-tuning of the initial condition 
or early matter domination~\cite{Kawasaki:1995vt}, 
to avoid the overabundance of the axion.

The axion potential is given by 
\begin{align}
 V_a = -\LQCD^4 \cos\left(\ol{\theta}(\phi) \right) + \Delta V. 
\end{align}
Here, $\Delta V := V_\CW|_{x = x_0}$ 
is the axion-dependent part of the CW potential which can spoil the axion solution. 
The shift of the $\theta$-angle due to $\Delta V$ is estimated as 
\begin{align}
 \Delta \theta 
= \frac{1}{\LQCD^4} 
 \left( \frac{\partial \Delta V}{\partial \phi} \right)   
\simeq 
     \frac{ m_0^2 M_Q^2}{8\pi^2 \LQCD^4} x_0^k \eps^{4+N}  \Omega
      \sin{\theta_0},   
\end{align}
with 
\begin{align}
    \Omega := \left( \beta_1 \frac{kN-4\pi x_0}{kN-2\pi x_0} -\beta_2 \right) 
      \left(\log\frac{m^2_0+m_{Q_2}^2}{e\mu^2} 
    + \frac{m_{Q_2}^2}{m_0^2}\log\left(1+\frac{m^2_0}{m_{Q_2}^2}\right)\right),   
\end{align}
where $\eps := e^{-\pi x_0/N}$. In our model, $k=12$ and $N=3$. 
Here, $m^2_{Q_2}$ is the mass squared of $Q_2$ defined in Eq.~\eqref{eq-mQi} 
with taking $x=x_0$ and dropping the negligible corrections at $\order{q}$. 
$\beta_{t}$ is the ratio $c_1/c_0$ in the expansion of the modular forms 
$\Ykr{k}{1_t}$ in Eq.~\eqref{eq-qexpand}, 
which is given by $\beta_1 = 472$ and $\beta_2 = 224$ 
for $\Ykr{12}{1_1}$ and $\Ykr{12}{1_2}$, respectively. 
Note that the $\order{q}$ correction proportional to $m_{Q_1}^2$ 
is canceled at the minimum $x=x_0$ and thus 
the leading term is $\eps^{4+N}$ which is proportional to 
$m_{Q_2}^2 \sim \order{\eps^{4}x_0^k M_Q^2}$. 
This shift of the angle should be less than $\order{10^{-10}}$ 
to be consistent with the measurement of the neutron electric dipole moment~\cite{Abel:2020pzs}.
A possible modulus $\tau$-dependence of $\LQCD$   
from the gauge kinetic function will also be suppressed by $\order{q}$, 
and thus is sub-dominant for the stabilization. 
Similarly the $\phi$ dependence of $\partial_\phi \ol{\theta}$ 
is also negligible. 
In the $A_4$ model discussed in the previous section, 
we obtain
\begin{align}
 \abs{\Delta \theta} \simeq  2\times 10^{-10} \times  \sin\theta_0 
\left(\frac{\abs{\Omega}}{10^4}\right)  
\left(\frac{m_0}{10^7~\GeV}\right)^2 
\left(\frac{M_Q}{M_p}\right)^2  
\left(\frac{x_0}{28}\right)^{12} 
\left(\frac{\eps}{10^{-12}}\right)^7. 
\end{align}
Thus the axion quality is ensured if $\eps \lesssim 10^{-12}$, 
where $\ita \sim 13$ as shown in Fig.~\ref{fig_VCW}. 
The required tuning becomes mild if $\sin \theta_0 \ll 1$ 
which might happen when the (generalized) CP is conserved at some level 
as discussed in e.g. Ref.~\cite{Feruglio:2023uof}~\footnote{See Refs.~\cite{Kobayashi:2020oji,Ahn:2023iqa} for recent studies on the strong CP problem relevant to the modular symmetry.}. 
In such a case, the strong CP problem is partially solved 
by the modular axion $\phi \sim \rta$.

\begin{figure}[t]
\centering
\includegraphics[width=0.65\hsize]{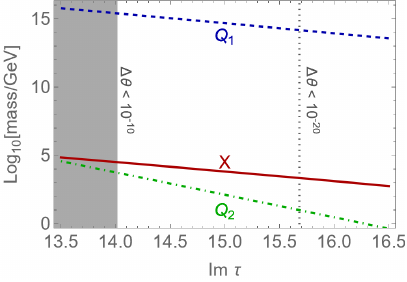}
\caption{\label{fig_mass}
Masses of the modulus $X\sim \ita$ (red) and vector-like quarks $Q_{1}$ (blue dashed) 
and $Q_2$ (green dot-dashed).   
In this plot, $m_0 = 10^{7}~\GeV$, $M_Q = 10^4\times \mu = M_p$, 
and $h=3$. 
}
\end{figure}

Before closing, we discuss the masses of the modulus $X\propto\ita$
and the vector-like quarks $Q_{1,2}$.  
The modulus mass may need to be heavier than $\order{10~\TeV}$,  
so that the modulus decay occurs before the big-bang nucleosynthesis 
and the moduli problem is avoided~\cite{Coughlan:1983ci,Goncharov:1984qm,Ellis:1986zt,deCarlos:1993wie}.   
After canonically normalizing the kinetic term Eq.~\eqref{eq-modkin}, 
the modulus mass is given by 
\begin{align}
\label{eq-mX2}
 m_X^2 = \frac{m_0^2 \tM_Q^2}{8\pi^2 h M_p^2} x_0^k \eps^2 
      (k-px_0)^2.   
\end{align}
Hence the modulus mass is related to $\Delta \theta$ as 
\begin{align}
\label{eq-mXQ}
 m_X \simeq&\ 
   c_0 \abs{k-px_0} \sqrt{\frac{\Delta \theta}{h \Omega \sin{\theta_0}}} 
      \frac{\LQCD^2}{\eps^{1+N/2} M_p} \\ \notag 
     \sim&\ 42~\TeV\times 
         \frac{c_0\abs{k -2\pi x_0/N}}{\sqrt{h \sin{\theta_0}}}  
         \left(\frac{\Delta\theta}{10^{-10}}\right)^{1/2}
         \left(\frac{10^4}{\Omega}\right)^{1/2}
         \left(\frac{\LQCD}{100~\MeV}\right)^2 
         \left(\frac{10^{-12}}{\eps}\right)^{5/2},    
\end{align}
in the $A_4$ model.
Thus the modulus mass decreases 
as the axion quality is higher by $\eps \to 0$. 
For $\Delta\theta/\sin\theta_0 \sim 10^{-10}$, 
$\eps \lesssim 10^{-12}$ is required to avoid the moduli problem~\footnote{
The lighter modulus would not be a problem 
if there is the thermal inflation~\cite{Yamamoto:1985mb,Lazarides:1985ja,Lyth:1995ka}, 
large Hubble induced mass during the inflation exists~\cite{Linde:1996cx,Nakayama:2011wqa}, 
and/or an inflation scale is lower than the modulus mass. 
The amount of relativistic axions produced from modulus decays, 
which may affect to $\Delta N_\mathrm{eff}$ 
is discussed in Refs.~~\cite{Cicoli:2012aq,Higaki:2012ar,Higaki:2013lra}.  
}.

The mass of the lighter vector-like quark $Q_2$, 
whose charge under $\intZ^T_3$ is $2$, is given by 
\begin{align}
 m_{Q_2} \sim \tM_Q x_0^{k/2} \eps^2 
            \sim 1~\TeV \times 
            \left(\frac{\tM_Q}{M_p}\right)    
            \left(\frac{x_0}{28}\right)^6 
            \left(\frac{\eps}{10^{-12}}\right)^2,   
\end{align}
thus $M_Q \sim M_p$ 
is necessary for $m_{Q_2} \gtrsim \order{\TeV}$ 
to be consistent with the LHC constraints~\cite{ATLAS:2020fgf,ATLAS:2022hnn,ATLAS:2022ozf,CMS:2020ttz,CMS:2021roc}. 
Conversely, 
if we require $m_{Q_2} \gtrsim \TeV$ and $\tM_Q \lesssim M_p$, 
the hierarchy parameter is bounded from below as 
$\eps \gtrsim 10^{-8}\times x_0^{-k/4}$, 
and thus larger weight is preferred to allow $\eps \sim \order{10^{-12}}$ 
without having the too light vector-like quark. 
Figure~\ref{fig_mass} shows the masses of the modulus $X\propto \ita$ 
and the vector-like quarks. 
As $\ita$ increases, the axion quality becomes better, 
whereas the modulus and vector-like quark masses decrease, 
and hence there is the $lower$ bound $\Delta\theta \gtrsim 10^{-13}$ 
for sufficiently heavy vector-like quark $Q_2$.

The vector-like quarks should be able to decay into SM particles. 
For illustration, if the vector-like quarks have the same quantum number 
as the down-type quarks, denoted by $d$, we can write down the superpotential
\begin{align}
\label{eq-Wmix}
 W_{\mathrm{mix}} = \sum_{i=1,2} \la_i \Ykr{14}{1_1} H_d q Q_i,   
\end{align}
where $H_d$ and $q$ are respectively the down-type Higgs doublet 
and the doublet SM quark. 
Here, the representations and weights of quarks are set to 
\begin{align}
\label{eq-SampleAssign}
& r_{q} = r_{d} = r_{\ol{Q}_1} = 1_0, \quad 
  r_{Q_{1,2}}  = r_{\ol{Q}_2} = 1_2,   
\quad 
k_{q} = h-k_{d} = 8, \quad k_{Q_{1,2}}-h = k_{\ol{Q}_{1,2}} = 6,
\end{align}
and those of $H_d$ is the trivial-singlet with weight $0$. 
We omit the flavor indices of the SM quarks. 
With this assignment, the interactions in Eqs.~\eqref{eq-super} and~\eqref{eq-Wmix}, 
as well as the SM Yukawa coupling $y H_d q d$ with $y$ being a constant, 
are realized, 
while the $\ol{Q}_i d$ is vanishing because of the negative weight.
The size of mixing $s_{Q_i}$ is estimated as 
\begin{align}
 s_{Q_i} \sim&\ 
\frac{\la_i x_0^{7} \Ykr{14}{1_1}  \vev{H_d}}
    {M_{Q} x_0^6 \Ykr{12}{1_i}} 
         \sim   \frac{\la_i x_0 \vev{H_d}}{M_{Q} \eps^{i-1}}   \notag  \\ 
         \sim&\ 10^{-3} \times \la_2 \left(\frac{x}{28}\right)  
                    \left(\frac{\vev{H_d}}{100~\GeV}\right)
                    \left(\frac{M_p}{M_{Q}}\right)
                    \left(\frac{10^{-12}}{\eps}\right), 
\end{align}
where $i=2$ in the second line, and we used $\Ykr{k}{1_t} \sim \eps^t$.
Thus the vector-like quarks decay promptly at the collider scale. 
It is noted that the CW potential induced by the mixing term 
is negligible because $\vev{H_d} \ll M_Q$.

\section{Summary and discussions} 
\label{sec-concl}

In this work, 
we point out that the modulus of the finite modular symmetry 
can be stabilized by the Coleman-Weinberg (CW) potential. 
For illustration, we studied the model with $\Gamma_3 \simeq A_4$ symmetry 
and two vector-like quark pairs which transform as non-trivial singlets, 
namely $1_1$ and $1_2$. 
This is the minimal possibility to cancel the mixed QCD anomaly 
of the finite modular symmetry, 
but to induce that of the $U(1)_\PQ$ symmetry. 
The CW potential has the global minimum at $\ita \gg 1$ 
if the corresponding modular form is non-trivial singlet under $A_4$, 
where the residual symmetry $\intZ^T_3$ remains unbroken. 
Since the potential along the $\rta$ direction is extremely flat 
due to this residual symmetry, 
we can regard $\rta$ as the QCD axion to solve the strong CP problem. 
The accidental $U(1)_\PQ$ arises due to the $\intZ^T_3$ symmetry 
and hence the PQ breaking effects are controlled by $\Gamma_3$ 
which is a different situation argued in Ref.~\cite{Heidenreich:2023pbi}.
We examined the condition to ensure the quality of this finite modular axion, 
and correlate with the masses of the modulus $X\sim \ita$
 and vector-like quarks. 
Interestingly, the modulus $X$ and the lighter quark $Q_2$
are expected to be $\order{\TeV}$ scale for $\Delta \theta < 10^{-10}$ 
and thus could be probed by cosmology and collider experiments.  
This mechanism can be generalized to other modular forms  
as long as the CW potential is dominated by that with non-zero $\intZ^T_3$-charge 
and the potential converges to zero for $\ita \gg 1$.

The modular $A_4$ symmetry in our scenario can not be used to explain the hierarchical structure of the SM fermions, 
as studied in Refs.~\cite{Petcov:2022fjf,Petcov:2023vws,Kikuchi:2023jap, Abe:2023ilq,Kikuchi:2023cap, Feruglio:2021dte,Novichkov:2021evw,Abe:2023qmr,Abe:2023dvr,deMedeirosVarzielas:2023crv, Kikuchi:2023dow,Kikuchi:2023fpl}, 
since the hierarchy parameter $\eps$ is $\order{10^{-12}}$ and is too small.
Such a tiny $\eps$ is required 
due to the correlation of the axion quality $\Delta \theta$ 
with the modulus mass in Eq.~\eqref{eq-mXQ}. 
This relation can be relaxed 
if the $\tau$-independent angle $\theta_0$ is small or the level $N$ is large. 
In the latter case, there exists a small hierarchy $|e^{2 \pi i \tau/N}|$ in the SM sector and a large one $|e^{2 \pi i \tau}|$ in the vector-like quark sector.\footnote{
In general,
for example, if one could break the periodicity $\tau \sim \tau + 1$ 
to $\tau \sim \tau + 1/n$ ($n>1$) in the vector-like $(Q,\bar Q)$ sector with keeping the periodicity 
$\tau \sim \tau +1$ in the SM fermion sector, then modular forms may appear 
$Y^{k}_r(\tau)$ in the SM fermion sector and  $Y^{k'}_{r'}(n\tau)$ in the  
$(Q,\bar Q)$ sector, and one could realize both a small hierarchy $|e^{2 \pi i \tau}|$ in the SM sector
and a large one $|e^{2 \pi i n\tau}|$ in the vector-like quark pairs.
}.
Thus, we could construct a model which explains the SM fermion hierarchies 
by the modulus stabilized by the mechanism proposed in this paper.  
An explicit model building is subject of our future work.

We have proposed a new scenario to lead to a large hierarchy 
$e^{- 2\pi \ita} \sim \order{10^{-36}}$ 
and a light axion by stabilizing the modulus at 
$\ita \gg 1$.
Although we have applied it to solve the strong CP problem by 
assuming $Q$ and $\bar Q$ as vector-like matter fields with the QCD colors, our scenario would 
be useful to explain other large hierarchies,  
which would be required for  
the proton stability, tiny neutrino masses, CP/flavor violations, quintessence 
and so on,  
together with light axions 
by assuming $Q$ and $\bar Q$ as visible or hidden matter fields. 
We would study it elsewhere.

In this work, we assume that the soft scalar masses of the vector-like quarks 
are independent of the modulus $\tau$
and supersymmetry is dominantly broken by other sector, 
so that the supersymmetry breaking sector does not contribute
to the stabilization of the modulus $\tau$. 
In addition, 
there would be contributions from gravitational instantons~\cite{Giddings:1987cg} 
which would change the potential structure. 
Those effects and other dynamical effects including stabilization mechanisms of other moduli, 
and uplifting of the vacuum energy if necessary, 
may become important in complete models such as superstring, 
but beyond the scope of this paper.

\section*{Acknowledgement} 

We thank Y.Abe and N.Aso for helpful discussions in the early stage of this work.  
The work of J.K. is supported in part by the Institute for Basic Science (IBS-R018-D1). 
This work is supported in part by he Grant-in-Aid for Scientific Research from the
Ministry of Education, Science, Sports and Culture (MEXT), Japan 
No.\ JP22K03601~(T.H.) and JP23K03375~(T.K.).

{\small
\bibliography{ref_modular} 
\bibliographystyle{JHEP} 
}

\end{document}